\RequirePackage{amsmath}
\documentclass[runningheads]{llncs}
\usepackage[T1]{fontenc}
\usepackage{graphicx}

\usepackage{amsmath}
\usepackage{amsfonts}
\usepackage{booktabs}

\usepackage{hyperref}
\usepackage{color}

\begin{document}
\title{Region of Interest Detection in Melanocytic Skin Tumor Whole Slide Images - Nevus \& Melanoma}
%
%


\author{Yi Cui\inst{1}
\and
Yao Li\inst{2}
\and
Jayson R. Miedema\inst{3}
\and
Sharon N. Edmiston\inst{4}
\and
Sherif Farag\inst{3}
\and
J.S. Marron\inst{2,3}
\and
Nancy E. Thomas\inst{3}}

\authorrunning{Cui et al.}

%

\institute{Department of Economics, University of North Carolina at Chapel Hill, Chapel Hill, NC 27599, USA\\
\email{yicui@unc.edu} \\
\and
Department of Statistics \& Operations Research, University of North Carolina at Chapel Hill, Chapel Hill, NC 27599, USA\\
\email{\{yaoli,marron\}@email.unc.edu}
\and
School of Medicine, University of North Carolina at Chapel Hill, Chapel Hill, NC 27516, USA\\
\email{\{jayson\_miedema,sherif\_farag,nancy\_thomas\}@med.unc.edu}
\and
Lineberger Comprehensive Cancer Center, University of North Carolina at Chapel Hill, Chapel Hill, NC 27599, USA\\
\email{edmiston@ad.unc.edu}}

\maketitle              
\begin{abstract}
Automated region of interest detection in histopathological image analysis is a challenging and important topic with tremendous potential impact on clinical practice. The deep-learning methods used in computational pathology may help us to reduce costs and increase the speed and accuracy of cancer diagnosis. We started with the UNC Melanocytic Tumor Dataset cohort that contains 160 hematoxylin and eosin whole-slide images of primary melanomas (86) and nevi (74). We randomly assigned 80\% (134) as a training set and built an in-house deep-learning method to allow for classification, at the slide level, of nevi and melanomas. The proposed method performed well on the other 20\% (26) test dataset; the accuracy of the slide classification task was 92.3\% and our model also performed well in terms of predicting the region of interest annotated by the pathologists, showing excellent performance of our model on melanocytic skin tumors. Even though we tested the experiments on the skin tumor dataset, our work could also be extended to other medical image detection problems to benefit the clinical evaluation and diagnosis of different tumors.

\end{abstract}
\section{Introduction}
The American Cancer Society predicted that in 2022 an estimated 99,780 cases of invasive and 97,920 cases of in-situ melanoma would be newly diagnosed and 7,650 deaths would occur in the US~\cite{Siegel2022Cancer2022}. The state-of-the-art histopathologic diagnosis of a melanocytic tumor is based on a pathologist’s visual assessment of its hematoxylin and eosin (H\&E)-stained tissue sections. However, multiple studies have suggested high levels of diagnostic discordance among pathologists in interpreting melanocytic tumors~\cite{Brochez2002Inter-observerLesions,Duncan1993HistopathologicStudy,Siegel2014Cancer2014}. Correct diagnosis of primary melanoma is key for prompt surgical excision to prevent metastases and in identifying patients with primary melanoma who are eligible for systemic adjuvant therapies that can improve survival. Alternatively, overdiagnosis can lead to unnecessary procedures and treatment with toxic adjuvant therapies. We are applying deep-learning methods in computational pathology to determine if we can increase diagnostic accuracy, along with increasing speed and decreasing cost. Here, we examine methods for improving Region of Interest (ROI) detection in melanocytic skin tumor Whole Slide Images (WSIs), which is an important step toward the computational pathology of melanocytic tumors.

Traditionally, expert pathologists visually identify the annotate the potential or related regions for melanoma and nevus, and then take a close look to classify certain types. However, this process is time-consuming and the accuracy is also not satisfactory~\cite{Farmer1996DiscordancePathologists}. One potential solution may be the combination of high-quality histopathological images and AI technology. Histopathological images have long been utilized in treatment decisions and prognostics for cancer. For example, histopathological images are used to score tumor grade in breast cancer to predict outcomes for cancer cases or perform histologic pattern classification in lung adenocarcinoma, which is critical for determining tumor grade and treatment for patients~\cite{Farahmand2022DeepCancer,Xie2019DeepCancer}. AI technology, like deep learning-based predictors trained on annotated and non-annotated data, could be a potentially efficient technology to improve early detection~\cite{Ianni2020TailoredWorkload}, help pathologists diagnose tumors, and inform treatment decisions to potentially improve overall survival rates.

Recently, with the advancement of machine learning, especially deep learning, many researchers have developed various frameworks and Convolutional Neural Network (CNN) architectures, like ZefNet~\cite{MatthewD.ZeilerandRobFergus2014VisualizingNetworks}, Visual geometry group (VGG)~\cite{Simonyan2015VeryRecognition}, ResNet~\cite{He2016DeepRecognition}, DenseNet~\cite{Huang2017DenselyNetworks}, etc., to solve the biomedical image computing and classification problems in the field of computer vision and pathology~\cite{Farahmand2022DeepCancer,Xie2019DeepCancer}. The idea of transfer-learning from these frameworks is to use a network that has been trained on unrelated categories on a huge dataset, like Imagenet, and then transfer its knowledge to the small dataset. Besides these transfer-learning-based methods, there also exist some methods that do not use the pre-trained model. These models are trained only by training datasets to update all the CNN parameters. In other words, deep learning largely expands our methods to deal with prediction and classification problems in pathology, and the applications include tumor classification~\cite{Liu2017DetectingImages,Wei2019Pathologist-levelNetworks}, cancer analysis and prediction~\cite{Farahmand2022DeepCancer,Xie2019DeepCancer}, cancer treatment prediction~\cite{Braman2020DeepStudy,Mobadersany2018PredictingNetworks} and so on. Therefore, in the field of computational pathology, more and more researchers use these deep learning methods on medical images that include rich information and features. Some papers~\cite{Farahmand2022DeepCancer,Noorbakhsh2020DeepImages,Wei2019Pathologist-levelNetworks} try to use WSIs and show high performance and accuracy of their models on certain types of cancers like breast cancers and uterine cancers. Recently, there has been some literature~\cite{Lu2015AutomatedImages} analyzing skin cancer based on histopathological images. However, previous literature does not include the ROI detection of melanocytic skin tumors and only has limited accuracy in classification as well as identification among various tumors.

\begin{figure}
\centering
\includegraphics[width=0.8\textwidth]{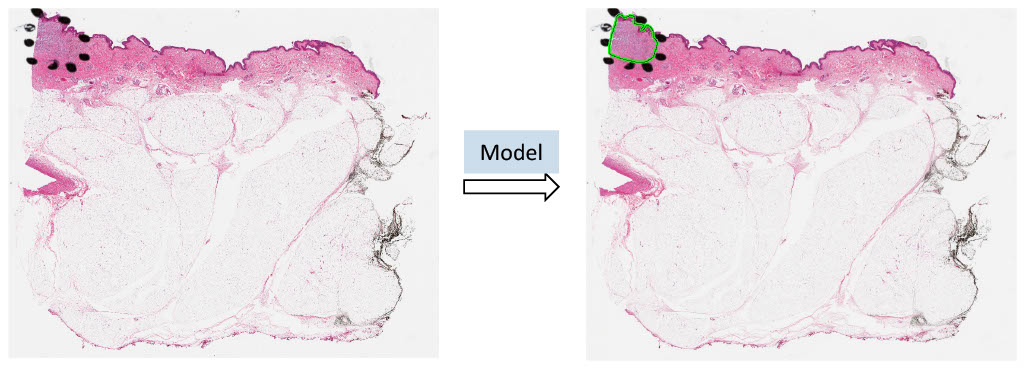}
\caption{ROI was annotated by black dots determined by pathologists. The predicted ROI was bounded by the green line on the right.} \label{f1}
\end{figure}

\begin{figure}
\centering
\includegraphics[width=1\textwidth]{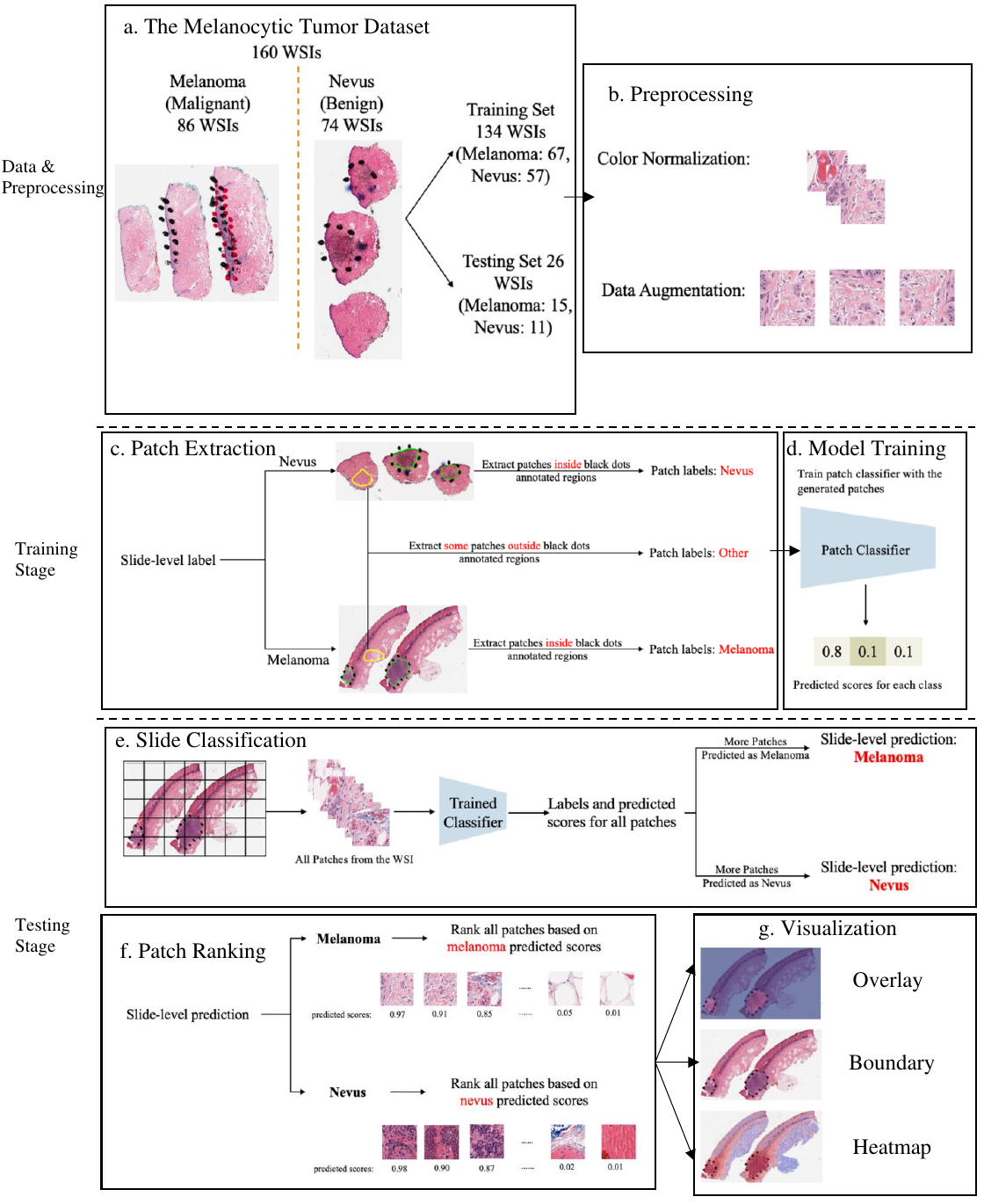}
\caption{Overview of the proposed detection framework. (a) The Melanocytic Tumor Dataset: Randomly assigned 80\% (134 WSIs) of data as the training set and 20\% (26 WSIs) of data as the testing set. (b) Preprocessing: color normalization~\cite{Macenko2009ANor} and data augmentation. (c) Extract melanoma, nevus and other patches from training data. (d) Model Training: Trained a 3-class patch classifier based on extracted patches. (e) Slide Classification: For each slide, generated predicted scores for all patches and calculated patch as well as slide classification accuracy. (f) Patch Ranking: Ranked all patches from a slide based on the corresponding predicted scores in the context of melanoma or nevus, depending on the slide classification result. (g) Visualization: Generated visualization results based on predicted scores.} \label{f2}
\end{figure}

Benefiting from the AI technology and histopathological images, we developed a deep neural network-based ROI detection method that could precisely detect the ROI in melanocytic skin tumors through WSIs and at the same time, classify the slides accurately. In Fig.~\ref{f1}, the slides had ROI indicated by black dots. Our goal was to automatically find this region without the use of black dots. The performance of our model could be seen as the green boundary in the right panel. Large images were broken into small patches and extracted abundant features from these patches~\cite{Cicek20163DAnnotation,Chen2018DeepLab:CRFs,Liu20183DVolumes,Milletari2016V-Net:Segmentation,Oktay2018AttentionPancreas}. In addition, we leveraged the partial information from annotations, also called “semi-supervised learning”, to enhance our model detection method. This method improved classification accuracy compared to previous approaches for certain kinds of tumors. Also, we proved our algorithm’s accuracy and robustness by decreasing our training samples to various subsets of the original training samples. Fig.~\ref{f2} illustrated the overview of our method. 

\section{Materials and Methods}
\subsection{Data}

\textit{Melanocytic Tumor Dataset cohort.} The melanocytic tumor dataset contained 86 melanomas (skin cancer) and 74 nevi (benign moles) WSIs. Besides slide-level labels, there were annotations made by pathologists on these slides. A slide might contain multiple slices of the same tissue, and pathologists annotated ROI on some slices for diagnosis purposes, but not others. We used Aperio ScanScope Console to scan the tissue samples with 20× magnification. 

\textit{Training Set.} We randomly selected 80\% (134 WSIs\footnote{It contained 71 melanoma (skin cancer) and 63 nevus (benign moles) WSIs.}) of data as our training set (Fig. 2a). For the training set, the slide-level labels (melanoma vs. nevus) are available, but the true annotations of ROI are not. While a portion of the ROIs in the slides were annotated, it should be noted that not all ROIs received annotation. This causes a challenge of using these annotations to evaluate the performance of the model in ROI detection task. However, we can still leverage these partial annotations to train a deep-learning model that can perform slide classification and ROI detection.

\textit{Testing Set.} We took the other 20\% (26 WSIs) as our testing set (Fig. 2a). For evaluation of our method and other baseline models, these 26 WSIs were manually annotated by our pathologist pathologists. Our model was trained on slides (from the training set) without ground-truth annotations with only partial information on WSIs. We used Aperio ImageScope Console to mark tumor boundaries as annotations and exported these annotations from the aforementioned software in the Extensible Markup Language (XML) format. It also included the annotated regions related to corresponding coordinates. We utilized these coordinates for each slide to figure out these regions solely from the rest of the image, labeled as melanoma or nevus.

\subsection{Data Preparation}
\textit{Data prep-processing: color normalization.} To minimize potential side effects of color recognition, we preprocessed all WSIs using previous color normalization methods~\cite{Ruifrok2001QuantificationDeconvolution}. There were different qualities or colors for scans performed in different labs or even the same lab for scans of the same WSIs processed at different times. The model may detect these undesirable changes to influence the feature extraction and even the following classification and ROI detection. Thus, we applied color normalization methods to these WSIs to ensure the slides that were processed under different circumstances were in the common, normalized space, which could enhance the robustness of model training and quantitative analysis (Fig. 2b). \textit{Data prep-processing: data augmentation.} First, tissue detection for the patches extracted from WSIs was completed. If we detected certain tissues, we would collect these tissues into patches and then finish the color normalization part. Data augmentation was then done by randomcorp, random horizontalflip and normalization of patches (Fig. 2b). And the edge features were restored accurately. \textit{Patch extraction.} Image slides were tiled into non-overlapping patches of 256 × 256 pixels in 20× magnification. Given a WSI, patches were extracted based on the slide-level label and annotations (Fig. 2c). If the slide-level label was nevus, all patches inside the annotated regions were labeled as nevus. If the slide-level label was melanoma, all patches inside the annotated regions were labeled as melanoma. Besides patches from annotated regions, some patches outside those regions were also extracted and labeled as \textit{other}. However, since not all ROIs were annotated by pathologists, there could be melanoma and nevus patches outside annotated regions. To avoid labeling those patches as \textit{other}, we manually extracted patches of other classes from regions. 

\subsection{Model training and assessment}
\begin{figure}
\begin{minipage}[b]{0.2\linewidth}
  \centering
  \centerline{\includegraphics[width=4.8cm]{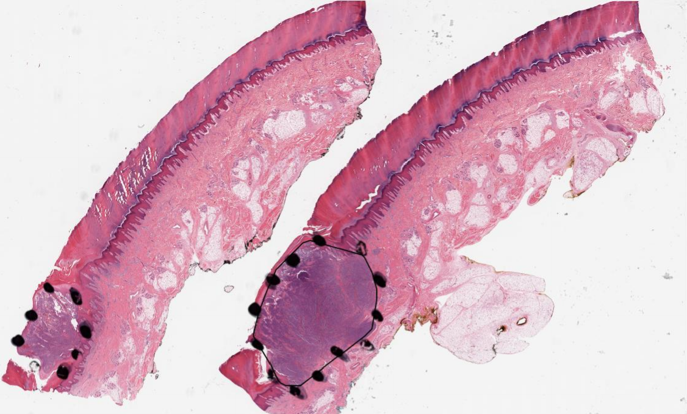}}
  \centerline{(a) Boundary} 
  \medskip
\end{minipage}
\hfill
\begin{minipage}[b]{.2\linewidth}
  \centering
  \centerline{\includegraphics[width=4.8cm]{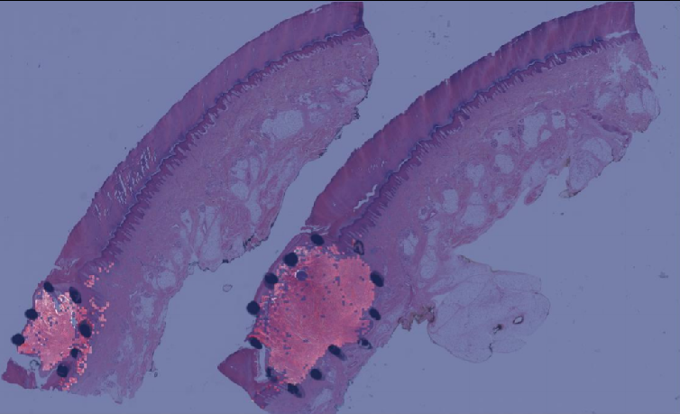}}
  \centerline{(b) Overlay}\medskip
\end{minipage}
\hfill
\begin{minipage}[b]{0.2\linewidth}
  \centering
  \centerline{\includegraphics[width=4.8cm]{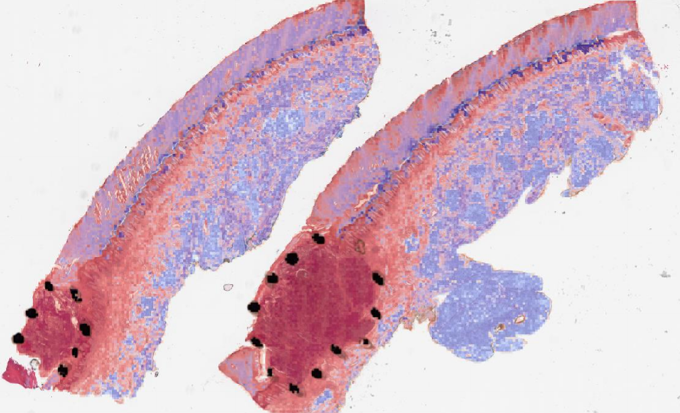}}
  \centerline{(c) Heatmap}\medskip
\end{minipage}

\begin{minipage}[b]{0.2\linewidth}
  \centering
  \centerline{\includegraphics[width=4.8cm]{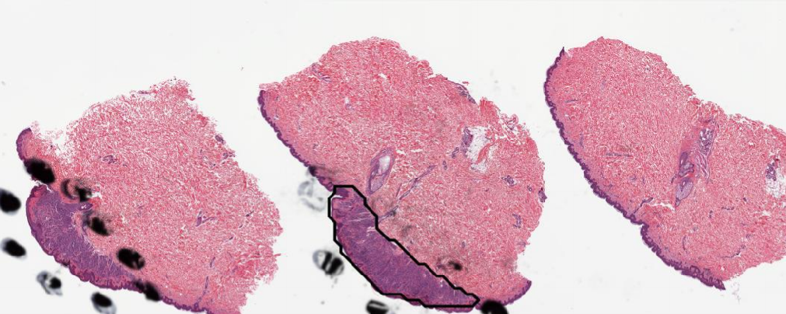}}
  \centerline{(d) Boundary}\medskip
\end{minipage}
\hfill
\begin{minipage}[b]{.2\linewidth}
  \centering
  \centerline{\includegraphics[width=4.8cm]{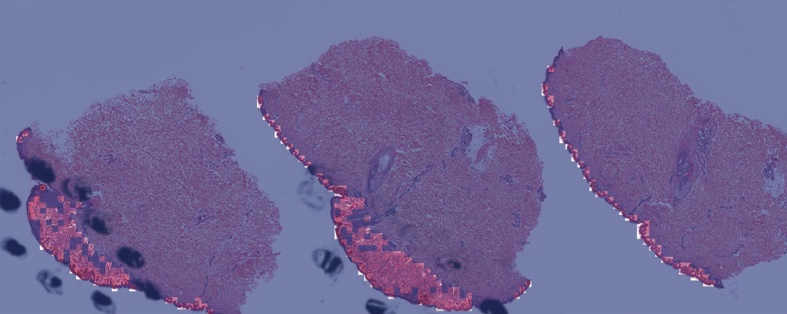}}
  \centerline{(e) Overlay}\medskip
\end{minipage}
\hfill
\begin{minipage}[b]{0.2\linewidth}
  \centering
  \centerline{\includegraphics[width=4.8cm]{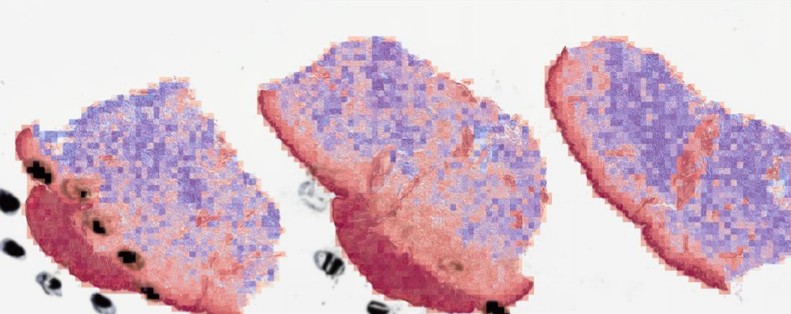}}
  \centerline{(f) Heatmap}\medskip
\end{minipage}
\caption{Visualization results for a {\it melanoma} sample and a {\it nevus} sample.}
\label{f4}
\end{figure}

\textit{Training patch classifier.} A three-class patch classification model (PCLA-3C) was trained on the labeled patches with VGG16~\cite{Simonyan2015VeryRecognition} as base architecture (Fig. 2d). Models were trained using this CNN architecture and by backpropagation, we manually changed the last layer’s parameters to optimize the model. The patch classifier would return a WSI with three key scores, corresponding to three categories (melanoma, nevus and other). \textit{Slide classification and ROI detection.} In the testing stage, all patches from a WSI were first fed into the trained patch classifier. Ignoring patches predicted as other, slide-level prediction was done by majority vote based on patches predicted as melanoma and nevus. If the number of patches labeled as melanoma exceeded the number of patches labeled as nevus in one WSI, we classified it as melanoma, and vice versa (Fig. 2e). For a WSI classified as melanoma, all the patches from this slide will be ranked by melanoma predicted scores. Otherwise, all the patches will be ranked by nevus predicted scores (Fig. 2f). \textit{Model assessment.} To evaluate the performance of ROI detection, the annotated ratio was measured to calculate Intersection over Union (IoU) for each slide. Given a slide, annotated ratio $\beta$ was calculated by the number of patches in the annotated region divided by the number of patches extracted from the slide: $\beta = \frac{A_p}{C_p}$,where $A_p$ is the number of patches in A (annotated region) and $C_p$ is the number of patches in C (WSI). Then, the top $n\beta$ patches based on predicted scores were classified as ROI, where $n$ was the total number of patches from a slide. For example, if  $\beta=0.2$ for a slide in the testing set, it means that 20\% of the regions in the slide are ROIs. Then, the model will predict the top 20\% of patches (based on the predicted scores) as patches in the ROIs. The performance was measured by Intersection over Union (IoU), which compared the annotated region and predicted ROI region. Since the framework was patch-based, IoU was calculated by the number of patches in the intersection region (the region in both annotated and predicted regions) divided by the number of patches in the union of the annotated and predicted ROI regions: $\text{IoU} = \frac{\underline{AB}_p}{\overline{AB}_p}$, where $\underline{AB}_p$ shows the number of patches in the region of $(A\cap B)$ and $\overline{AB}_p$ shows the number of patches in the region of $(A\cup B)$. A is annotated region and B is the predicted/highlighted region. \textit{Visualization.} The detection methods could provide three types of visualization maps: boundary, overlap and heatmap (examples were in Fig.~\ref{f4}). Three visualization maps will be generated based on the predicted scores calculated in the ROI detection section (Fig. 2g). The overlap map highlighted top-ranked patches in a WSI and masks other areas with a transparent blue color (Fig. 3a, 3d). The percentage of highlighted patches equaled $\beta$ (the annotated ratio). Therefore, the highlighted region was also the predicted ROI. The boundary map showed the boundary of the largest ROI cluster based on the highlighted patches, where the highlighted patches were clustered by OPTICS algorithm~\cite{Ankerst1999OPTICS:Structure} (Fig. 3b, 3e). The last one was a heatmap where red covered regions that had high predicted scores and blue covered regions that had low predicted scores (Fig. 3c, 3f).

\section{Results}
\subsection{Method Comparison}
Two methods were tested on the melanocytic skin tumor dataset to do ROI detection and slide classification: 1) CLAM (clustering-constrained attention multiple instance learning)~\cite{Lu2021Data-efficientImages}, 2) PCLA-3C (the proposed patch-based classification model). The 160 WSIs from UNC Melanocytic Tumor Dataset cohort were randomly split into training and testing sets with 134 for training and 26 for testing. Both methods were trained on the training set, and the performances on both training and testing sets were evaluated. Visualization results and code could be found on GitHub\footnote{\href{https://github.com/cyMichael/ROI_Detection}{https://github.com/cyMichael/ROI\_Detection}}.

\textit{Computational configuration.} All analyses were used by Python. Images were analyzed and processed using OpenSlide. All the computational tasks were finished on UNC Longleaf Cluster with Linux (Tested on Ubuntu 18.04) and NVIDIA GPU (Tested on Nvidia GeForce RTX 3090 on local workstations). NVIDIA GPUs supports were followed to set up and configure CUDA (Tested on CUDA 11.3) and the torch version should be greater than or equal to 1.7.1.

\subsection{Model Validation and Robustness}
We trained the model based on different proportions of the training dataset, but the results were based on the testing set (26 WSIs), see table~\ref{t1}. There was a high agreement between the predictions of the ROI by PCLA-3C and the true ones, showing the accuracy of our automatic ROI detection.

\begin{table}
\centering
\caption{Table captions should be placed above the
tables.}\label{t1}
\begin{tabular}{|l|l|l|}
\toprule
Evaluation metrics &  PCLA-3C & CLAM\\
\midrule
Patch classification accuracy &  0.892 & -\\
classification accuracy &  0.923 & 0.692\\
IoU & 0.382 & 0.112\\
\bottomrule
\end{tabular}
\end{table}

By using the training data, our method achieved an accuracy of 92.3\% in slide-level classification and IoU rate of 38.2\% in the ROI detection task on the testing set. Our method achieved better accuracy than CLAM with an accuracy of 69.2\% in slide-level classification and IoU rate of 11.2\% in the ROI detection task. Also, we analyzed the robustness results in the supplementary information, showing the accuracy was 0.7866 (95\% CI, 0.761-0.813) at the patch level, and accuracy was 0.885 (95\% CI, 0.857-0.914) at the slide level by using 80\% (107 WSIs) of the original training set. Our true testing data were kept unchanged since these data included true annotations. However, the training data did not include the true annotations. As in the PCLA-3C, the improvements in patch classification accuracy, slide classification accuracy and IoU showed the importance of annotations in the training of deep learning classifiers for prediction. Also, we showed that patch classification results can be used to predict the slide-level label accurately. This is important as accurate tumor type is the clinical biomarker for future treatment. In summary, our deep-learning-based framework has outperformed the state-of-the-art ROI detection method~\cite{Lu2021Data-efficientImages}, leading to better model visualization and interpretation. This is quite crucial in medical imaging fields and related treatment recommendations.

\subsection{Misclassified Slides Discussion}
\begin{figure}
\begin{minipage}[b]{1.0\linewidth}
  \centering
  \centerline{\includegraphics[width=5cm]{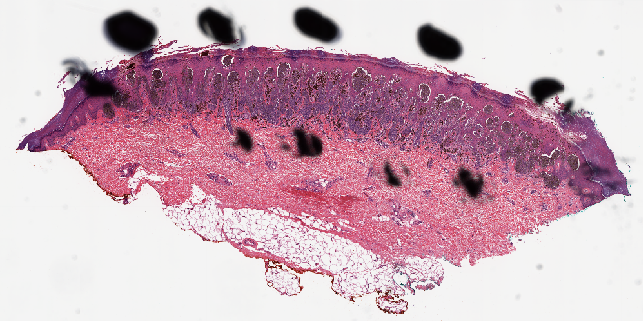}}
  \centerline{(a) Original WSI}\medskip
\end{minipage}
\begin{minipage}[b]{.48\linewidth}
  \centering
  \centerline{\includegraphics[width=5cm]{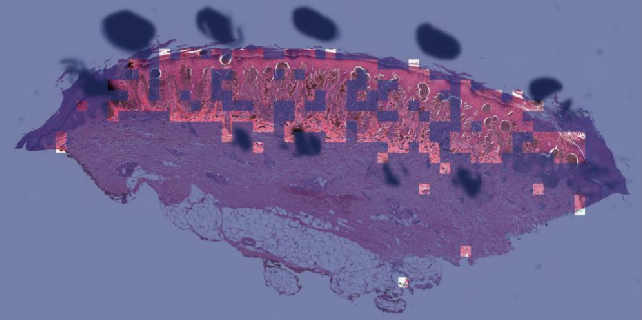}}
  \centerline{(b) Overlay}\medskip
\end{minipage}
\hfill
\begin{minipage}[b]{0.48\linewidth}
  \centering
  \centerline{\includegraphics[width=5cm]{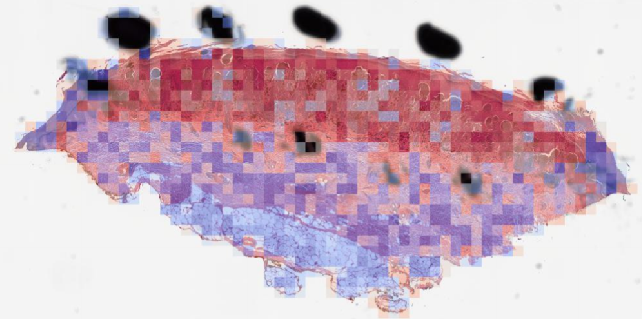}}
  \centerline{(c) Heatmap}\medskip
\end{minipage}
\vspace{-10pt}
\caption{Visualization results for a misclassified case 1.}
\label{f6}
\vspace{-10pt}
\end{figure}

\begin{figure}
\begin{minipage}[b]{1.0\linewidth}
  \centering
  \centerline{\includegraphics[width=5cm]{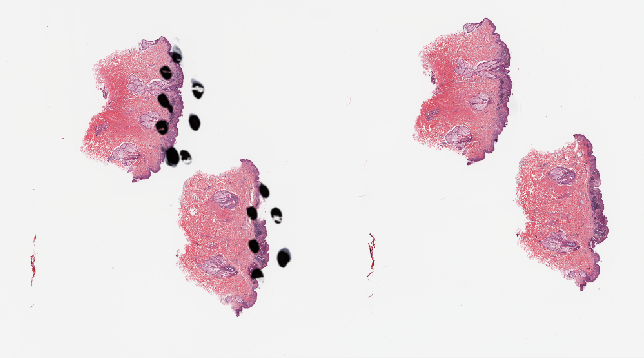}}
  \centerline{(a) Original WSI}\medskip
\end{minipage}
\begin{minipage}[b]{.48\linewidth}
  \centering
  \centerline{\includegraphics[width=5cm]{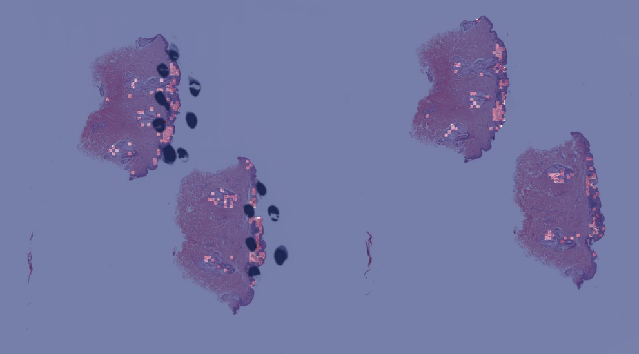}}
  \centerline{(b) Overlay}\medskip
\end{minipage}
\hfill
\begin{minipage}[b]{0.48\linewidth}
  \centering
  \centerline{\includegraphics[width=5cm]{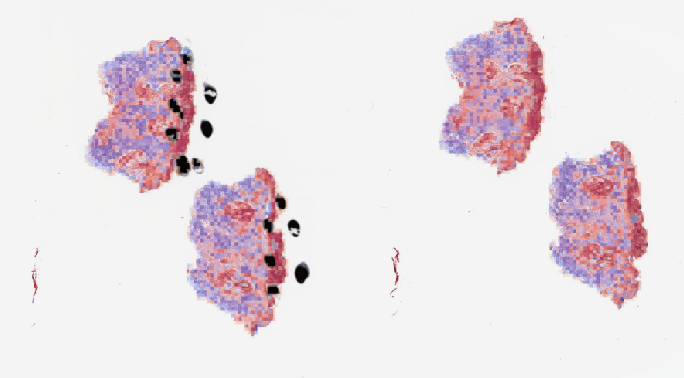}}
  \centerline{(c) Heatmap}\medskip
\end{minipage}
\vspace{-10pt}
\caption{Visualization results for a misclassified case 2.}
\label{f7}
\vspace{-10pt}
\end{figure}

The proposed method PCLA-3C only misclassified two slides in the testing set. The two WSIs are both labeled as nevus but misclassified as melanoma by the model (see the two slides and corresponding visualization results in Fig.~\ref{f6} and Fig.~\ref{f7}). The slide in Fig.~\ref{f6} is not a typical nevus and it has the features of a pigmented spindle cell nevus, which is one diagnostic challenge of melanocytic skin tumor. However, the slide in Fig.~\ref{f7} is a routine type of nevus. The reason that PCLA-3C misclassified the slide could be based on the difference in color. In general, the ROIs in melanoma cases were dark, while those in nevus cases were light. As shown in Fig. 7b, there were some dark areas outside the annotated ROIs, which contributed to the misclassification of slides and the incorrect detection of ROIs.

\section{Discussion}
In this work, we presented deep-learning-based classifiers for predicting the correct tumor types with and without annotations. Using high-quality WSIs from the UNC Melanocytic Tumor Dataset cohort annotated by our pathologists, we systematically selected the proper cases for training and testing. Heatmap, boundary and overlay figures exerted by PCLA-3C showed a considerable agreement with annotations finished by our pathologist group. Also, as shown in table~\ref{t1} and table~\ref{t2}, the test results showed that PCLA-3C had higher accuracy in patch level, slide level and ROI level by just using limited WSIs as the training set than CLAM.

\begin{table}
  \caption{Robustness performance of patch classification accuracy, slide classification and IoU by PCLA-3C and CLAM using different splits of the original training set. Since CLAM does not do patch classification, it does not have patch classification accuracy.}
  \label{t2}
  \centering
  \resizebox{12.2cm}{!}{
  \begin{tabular}{c||cc|cc||cc|cc}
    \toprule
     & \multicolumn{4}{c||}{20\% split}  & \multicolumn{4}{c}{40\% split}               \\ \cmidrule(r){2-9}
     & \multicolumn{2}{c|}{PCLA-3C}  & \multicolumn{2}{c||}{CLAM} & \multicolumn{2}{c|}{PCLA-3C}  & \multicolumn{2}{c}{CLAM}               \\
    \midrule
             & Mean     & 95\% CI & Mean     & 95\% CI  & Mean     & 95\% CI & Mean     & 95\% CI \\
    \cmidrule(r){2-9}
    Patch classification accuracy& 0.6397&	[0.5193, 0.7601]& - & - & 0.7887&	[0.7536, 0.8238] & - & -\\
    Slide classification accuracy     & 0.7406&	[0.6627, 0.8185] & 0.6710&	[0.6386, 0.7033]&0.8430&	[0.8043, 0.8817] & 0.6976&	[0.6619, 0.7333] \\
    Intersection over Union     & 0.3026&	[0.2394, 0.3327]&	0.0427 &	[0.0342, 0.0512] &  0.3402&	[0.3057, 0.3784] & 0.0524&	[0.0297, 0.0751] \\
    \midrule
    & \multicolumn{4}{c||}{60\% split}  & \multicolumn{4}{c}{80\% split} \\ \cmidrule(r){2-9}
     & \multicolumn{2}{c|}{PCLA-3C}  & \multicolumn{2}{c||}{CLAM} & \multicolumn{2}{c|}{PCLA-3C}  & \multicolumn{2}{c}{CLAM}               \\
    \midrule
    & Mean     & 95\% CI & Mean     & 95\% CI & Mean     & 95\% CI & Mean     & 95\% CI\\
    \cmidrule(r){2-9}
    Patch classification accuracy& 0.8191 & [0.7766, 0.8616]& - & - & 0.8210& [0.7949, 0.8471] & - & - \\
    Slide classification accuracy     & 0.8721&	[0.8458, 0.8985]& 0.7097&	[0.6830, 0.7364]& 	0.8885&	[0.8607, 0.9163] & 0.7258&	[0.7117, 0.7399] \\
    Intersection over Union     & 0.3652&	[0.3369, 0.3934]& 0.0621&	[0.0428, 0.0814] & 0.3710	&[0.3335, 0.4084] & 0.1103	&[0.0529, 0.1677]\\
    \bottomrule
  \end{tabular}}
  \vspace{-0.5cm}
\end{table}

Some recent studies have also examined tumors by using the deep-learning architecture in the medical imaging field. Most literature mainly studied the effects of CNN-based methods on different cancers like breast cancer and skin cancer, and achieved high accuracy on the classification task. Khalid et al.~\cite{Hosny2020SkinLearning} have utilized deep learning and transfer learning to classify skin cancers. Some literature~\cite{Liu2022AImages,Murtaza2020DeepChallenges} tried to solve the classification problem in breast cancer by deep learning methods. Besides, Farahmand et al.~\cite{Farahmand2022DeepCancer} have not only classified the WSI accurately, but they are also focused on the ROI detection tasks and achieved nice results. From Lu et al.~\cite{Lu2021Data-efficientImages}, CLAM has been used to solve the detection of renal cell carcinoma and lung cancer. CLAM is proposed to do slide classification and ROI detection, which does not require pixel or patch-level labels. However, when applied to the melanocytic skin tumor dataset, the ROI detection of this method is not satisfactory. Lerousseau et al.~\cite{Lerousseau2020WeaklySegmentation} introduced a weakly supervised framework (WMIL) for WSI segmentation that relies on slide-level labels. Pseudo labels for patches were generated during training based on predicted scores. Their proposed framework has been evaluated on multi-locations and multi-centric public data, which demonstrated a potentially promising approach for us to further study the WSIs. 

Here we reported on a novel method that performed automated ROI detection on primary skin cancer WSIs. It improved the performance of the state-of-the-art method by a large margin. 

In most places, the diagnostic pathologists will manually scan all the slides to analyze the tumor types. Thus, it is convenient and cheap to apply the deep-learning method to these existing WSIs. The high accuracy of our deep learning-based method results has made huge progress toward digital assistance in diagnosis.

The key strength of our model is that it overcomes the lack of ground-truth labels for the detection task. The performance of previous methods was not satisfactory on melanocytic WSIs. One reason is that melanocytic tumors are difficult to diagnose and detect, and the literature reports 25–26\% of discordance between individual pathologists for classifying a benign nevus versus malignant melanoma~\cite{Hekler2019Pathologist-levelNetworks}. Using only slide-level labels was hard to train a promising method. The success of our method means the combination of partial information from annotations and patch-level information could largely enhance the analysis of melanocytic skin tumors.

The weakness of our model is that our model does not classify all the WSIs accurately. Our slide classification is 92.3\%, so we could not rely completely on the model (PCLA-3C). Two WSIs (true label: nevus) in the testing set were misclassified as melanoma. Although our method does not perform the same as the gold standard, our results can assist pathologists in efficiently classifying the WSIs and finding the ROI.

In summary, the deep-learning architecture that we developed and utilized in this study could produce a highly accurate and robust approach to detect skin tumors and predict the exact type of tumors. Given that it takes lots of time to examine the patients’ WSIs, besides the conventional methods, our efficient AI method could help medical staff save time and improve the efficiency and accuracy of diagnosis, which benefits each patient in the future. We expect that our approach will be generalizable to other cancer-related types, not restricted to skin cancer, or breast cancer~\cite{Xie2019DeepCancer}, and vision-related treatment outcome predictions. The deep-learning-based framework could also be widely applied in identification and prediction in diagnostics. In the future, we plan to extract some detailed information from high-quality WSIs and then improve our model to get higher accuracy in detection and prediction. Future work will also include further improvements in the ROI detection performance by incorporating extra information into the model, such as gene expression and clinical data.

\begin{credits}
\subsubsection{\ackname} We would like to thank the School of Medicine of the University of North Carolina at Chapel Hill for providing study materials. All authors performed final approval of the paper, are accountable for all aspects of the work, confirm that we had full access to all the data in the study and accept responsibility to submit for publication. This study was supported by the National Cancer Institute at the National Institutes of Health P01CA206980, R01CA112243 grants and the UNC Health Foundation.


\subsubsection{\discintname}
Approval for the study was granted by the Institutional Review Board (IRB) of UNC Chapel Hill. The tissues and data for the 
Melanocytic Tumor Dataset cohort were retrieved under permission from IRB \# 22-0611. The IRB determined the research met the criteria for waiver of informed consent for research [45 CFR 46.116(d)] and waiver of HIPAA authorization [45 CFR 164.512(i)(2)(ii)]. The study complies with all regulations. The datasets used and/or analyzed during the current study are available from the corresponding author on reasonable request. All data generated or analyzed during this study are included in this paper~\cite{Conway2019IdentificationDiagnosis}.
\end{credits}
%
%
%
\bibliographystyle{splncs04}
\bibliography{mybibliography}

\end{document}